\documentclass{article}
\usepackage{ijcai23}
\usepackage{times}
\usepackage{soul}
\usepackage{xcolor}
\usepackage{url}
\usepackage[hidelinks]{hyperref}
\usepackage[utf8]{inputenc}
\usepackage[small]{caption}
\usepackage{graphicx}
\usepackage{amsmath}
\usepackage{amsthm}
\usepackage{amsfonts}
\usepackage{booktabs}
\usepackage{algorithm}
\usepackage{algorithmic}
\usepackage[switch]{lineno}

\usepackage{subfig}
\usepackage{multirow}
\usepackage{rotating}
\usepackage[thinlines]{easytable}

\newcommand{\Mod}[1]{\ (\mathrm{mod}\ #1)}

\newcommand{\projectname}{VERTICES}

\newtheorem{theorem}{Theorem}
\newtheorem{lemma}[theorem]{Lemma}

\title{VERTICES: Efficient Two-Party Vertical Federated Linear Model with TTP-aided Secret Sharing}

\author{
Mingxuan Fan$^1$
\and
Yilun Jin$^1$\and
Liu Yang$^{1,2}$\and
Zhenghang Ren$^1$\and
Kai Chen$^{1,2}$
\affiliations
$^1$iSING Lab, The Hong Kong University of Science and Technology\\
$^2$Clustar\\
\emails
mfanae@connect.ust.hk,
yilun.jin@connect.ust.hk,
lyangau@cse.ust.hk,
zrenak@cse.ust.hk,
kaichen@cse.ust.hk
}

\begin{document}

\maketitle

\begin{abstract}
    Vertical Federated Learning (VFL) has emerged as one of the most predominant approaches for secure collaborative machine learning where the training data is partitioned by features among multiple parties. Most VFL algorithms primarily rely on two fundamental privacy-preserving techniques: Homomorphic Encryption (HE) and secure Multi-Party Computation (MPC). Though generally considered with stronger privacy guarantees, existing general-purpose MPC frameworks suffer from expensive computation and communication overhead and are inefficient especially under VFL settings.
    This study centers around MPC-based VFL algorithms and presents a novel approach for two-party vertical federated linear models via an efficient secret sharing (SS) scheme with a trusted coordinator. Our approach can achieve significant acceleration of the training procedure in vertical federated linear models of between 2.5$\times$ and 6.6$\times$ than other state-of-the-art MPC frameworks  under the same security setting.
\end{abstract}

\section{Introduction}

Machine Learning has become an increasingly important and powerful technique in many applications under the rapid growth of data and information technology. Empowered by the large amount of data, this technique has been embraced by many companies or individual data owners to train a centralized model and is nowadays the basic building block of most intelligent systems. Nevertheless, privacy concerns arise under circumstances where the training data is scattered across different data holders. Simply exchanging sensitive data for collaborative training leads to privacy leakage which is widely considered a serious problem of data security, while training on local data limits the capability of traditional machine learning and results in the "isolated data island" phenomenon. 

To address such problems, Federated Learning (FL)~\cite{DBLP:journals/corr/abs-1902-04885} has emerged as a practical technique for collaborative privacy-preserving machine learning. Based on the different ways of splitting training data, FL can be further classified into Horizontal Federated Learning (HFL) where data is partitioned across different samples and Vertical Federated Learning (VFL) where data is partitioned across different features. VFL~\cite{yang2023survey} \cite{liu2022vertical} is fairly common nowadays in cross-silo situations, enabling companies to collaboratively train a shared model while possessing only partial features of common samples. Differential Privacy (DP)~\cite{10.1007/978-3-540-79228-4_1}, Homomorphic Encryption~\cite{Rivest1978ONDB} and secure Multi-Party Computation (MPC) are most extensively used as the foundational techniques for preserving data privacy during collaborative computations. 

Different from the other two approaches, MPC is capable of achieving provable privacy guarantees while maintaining the correctness of the calculated result without accuracy drop. Nevertheless, as one the of the most extensively used approach in MPC, SS suffers from heavy communication and computation cost for secret multiplications due to the costly preparation phase for randomness generation in the two-party setting, as each multiplication would consume a Beaver Multiplication Triple (MT) generated via either HE or OT in the preparation phase to prevent privacy leakage \cite{7958569}. In machine learning tasks with high-dimensional data where large-scale matrix multiplications are ubiquitous, this offline preparation phase can take orders of magnitude longer time than the online execution phase.

In an attempt to mitigate the expensive offline preparation cost of two-party SS frameworks, a semi-honest and non-colluding Trusted Third Party (TTP) is considered for assistance. Most existing solutions either rely on TTP-aided randomness generation~\cite{10.1007/978-3-540-88313-5_13} \cite{gascon2016privacy} \cite{10.1145/3196494.3196522} that offload this process to a TTP to avoid expensive OT and HE operations, or replicated secret sharing (RSS) schemes~\cite{10.1145/2976749.2978331} \cite{10.1145/3243734.3243760} \cite{wagh2021f} to get rid of the requirements of multiplication triples. However, both approaches are sub-optimal in terms of efficiency under VFL settings. The former fails to fully utilize the computation resources of all participants and introduces high communication overhead exchanging the randomness, while the latter introduces excessive computation complexity due to the increased number of matrix multiplications in VFL tasks.

Motivated by these limitations in existing approaches, we propose a novel and efficient secret sharing scheme with TTP assistance for two-party vertical federated linear models. We notice that on one hand, the expensive communication overhead in the former approach results from its inner dependency on the multiplication triples. On the other hand, RSS brings unnecessary and redundant computation overhead in two-party settings as each secret is shared in three parts. To overcome these obstacles, we design 1) a TTP-aided secret multiplication technique which no longer requires pre-generated multiplication triples and 2) an optimized sharing scheme with removable masks to achieve both communication and computation efficiencies over existing approaches. Assuming a semi-honest and non-colluding adversary model, our approach achieves between 2.5$\times$ and 6.6$\times$ acceleration of the training time compared with existing secret sharing frameworks under the same security settings.

\section{Preliminaries}

\subsection{Overview of Vertically Partitioned Machine Learning }

In the general two-party VFL setting, parties are distinguished into the an active party who holds the label of the data and partial features and a passive party who holds the remaining features. Without the loss of generality, we consider two party $\mathcal{A}$ and $\mathcal{B}$ where $\mathcal{A}$ is the passive party and $\mathcal{B}$ is the active party. Denote the dataset as $\mathcal{D}\{\mathbf{X}, \mathbf{Y}\}$ where $\mathbf{X} \in \mathbb{R}^{n \times d}$ and $\mathbf{Y} \in \mathbb{R}^{n \times 1}$. $\mathbf{X}$ is vertically partitioned into $\mathbf{X}_a \in \mathbb{R}^{n \times d_1}$ hold by $\mathcal{A}$ and $\mathbf{X}_b \in \mathbb{R}^{n \times d_2}$ hold by $\mathcal{B}$ where $d_1 + d_2 = d$. Note that before VFL setup, the participants need to perform Private Set Intersection (PSI)~\cite{pinkas2014faster} which provides them the corresponding IDs of the matched samples. For simplicity, we skip this part and focus on the training process.

In both linear regressions and logistic regressions, given the trainable weights $\mathbf{w}_a$ and $\mathbf{w}_b$ corresponding to the features of the training data $\mathbf{X}_a$ and $\mathbf{X}_b$, the loss $L$ in linear regression tasks is generally defined as \[L = || \mathbf{\hat{Y}} - \mathbf{Y}||^2 = || \mathbf{X}_a \mathbf{w}_a + \mathbf{X}_b \mathbf{w}_b - \mathbf{Y}||^2\] In logistic regression tasks, the loss $\mathbf{L}$ is generally defined as \[\mathbf{L} = -\mathbf{Y} \log(\mathbf{\hat{Y}}) - (1-\mathbf{Y}) \log(1 - \mathbf{\hat{Y}})\] where $\mathbf{\hat{Y}} = \sigma(\mathbf{X}_a \mathbf{w}_a + \mathbf{X}_b \mathbf{w}_a)$. The gradient $\mathbf{g}$ for both tasks can be calculated as $\displaystyle \mathbf{g} = \mathbf{X}^T(\mathbf{\hat{Y}} - \mathbf{Y})$.

As a non-linear activation function, the sigmoid function $\sigma(x)$ is hard to evaluate in secret sharing calculations. Previous literature has proposed many different approximation methods for the sigmoid function, including Piece-wise Approximation~\cite{7958569}, Taylor expansion~\cite{kim2018secure}, and minimax approximation~\cite{chen2018logistic}. In this work, we adopted degree-3 minimax approximation of $\sigma(x) = q_0 x^3 + q_1 x + q_2$ considering both the accuracy and efficiency of the activation function.

\subsection{Additive Secret Sharing}

In this section, we provide some basic notations in secret sharing. We denote the additive sharing of a $l$-bit variable $x$ in ring $\mathbb{Z}_{2^l}$ as $\langle x \rangle _0$, $\langle x \rangle _1$ such that $\langle x \rangle _0 + \langle x \rangle _1 \equiv x \Mod{2^l}$. To perform addition operations on shared values $\langle z \rangle = \langle x \rangle + \langle y \rangle$, each party $i \in \{0, 1\}$ locally computes $\langle z \rangle _i = \langle x \rangle _i + \langle y \rangle _i$. To perform multiplications of shared values $\langle z \rangle = \langle x \rangle \cdot \langle y \rangle$, each party $i \in \{0, 1\}$ needs to generate a shared multiplication triple  $\langle c \rangle = \langle a \rangle \cdot \langle b \rangle$ so that they can collaboratively compute $\langle z \rangle_i = i \cdot (x + a) \cdot (y + b) - (y + b) \cdot \langle a \rangle_i - (x + a) \cdot \langle b \rangle_i + \langle c \rangle_i$. Note that all these arithmetic operations are performed in ring $\mathbb{Z}_{2^l}$.

\begin{algorithm}[tb]
    \caption{Forward Calculation for Linear Regression}
    \label{alg:lin_fw}
    \textbf{Input}: Secret shares hold by party $\mathcal{A}$, $\mathcal{B}$ and coordinator $\mathcal{C}$ in the current context\\
    \textbf{Output}: Aggregated masked sum $\mathbf{S}$ hold by coordinator $\mathcal{C}$
    \begin{algorithmic}[1] 

        \STATE Party $\mathcal{A}$ calculates $\mathbf{F}_\mathcal{A} = \mathbf{X}_a \langle \mathbf{w}_a \rangle_0 + \langle \mathbf{X}_b \rangle_0 \langle \mathbf{w}_b \rangle_0  + \alpha_0$ and sends $\mathbf{F}_\mathcal{A}$ to party $\mathcal{C}$

        \STATE Party $\mathcal{B}$ calculates $\mathbf{F}_\mathcal{B} = \mathbf{X}_b \langle \mathbf{w}_b \rangle_1 + \langle \mathbf{X}_a \rangle_1 \langle \mathbf{w}_a \rangle_1 - \mathbf{Y} + \alpha_1$ and sends $\mathbf{F}_\mathcal{B}$ to party $\mathcal{C}$

        \STATE Party $\mathcal{C}$ calculates $\mathbf{F}_\mathcal{C} = \langle \mathbf{X}_a \rangle_0 \langle \mathbf{w}_a \rangle_1 + \langle \mathbf{X}_b \rangle_1 \langle \mathbf{w}_b \rangle_0$ and sums up $\mathbf{S} = \mathbf{\hat{Y}} - \mathbf{Y} + \alpha = \mathbf{F}_\mathcal{A} + \mathbf{F}_\mathcal{B} + \mathbf{F}_\mathcal{C} = \mathbf{X}_a \mathbf{w}_a + \mathbf{X}_b \mathbf{w}_b - \mathbf{Y} + \alpha$ 
        
    \end{algorithmic}
\end{algorithm}

\begin{algorithm}[tb]
    \caption{Forward Calculation for Logistic Regression}
    \label{alg:log_fw}
    \textbf{Input}: Secret shares hold by party $\mathcal{A}$, $\mathcal{B}$ and coordinator $\mathcal{C}$ in the current context; the random mask $\gamma$; minimax approximation coefficients $(q_0, q_1, q_2)$ \\
    \textbf{Output}: Aggregated masked sum $\mathbf{S}$ hold by coordinator $\mathcal{C}$
    \begin{algorithmic}[1] 

        \STATE Party $\mathcal{A}$ calculates $\mathbf{F}'_\mathcal{A} = \mathbf{X}_a \langle \mathbf{w}_a \rangle_0 + \langle \mathbf{X}_b \rangle_0 \langle \mathbf{w}_b \rangle_0$ and sends $\gamma \mathbf{F}'_\mathcal{A}$ to party $\mathcal{C}$

        \STATE Party $\mathcal{B}$ calculates $\mathbf{F}'_\mathcal{B} = \mathbf{X}_b \langle \mathbf{w}_b \rangle_1 + \langle \mathbf{X}_a \rangle_1 \langle \mathbf{w}_a \rangle_1$ and sends $\gamma \mathbf{F}'_\mathcal{B}$ to party $\mathcal{C}$

        \STATE Party $\mathcal{C}$ calculates $\mathbf{F}'_\mathcal{C} = \langle \gamma \mathbf{X}_a \rangle_0 \langle \mathbf{w}_a \rangle_1 + \langle \gamma \mathbf{X}_b \rangle_1 \langle \mathbf{w}_b \rangle_0$ and sums up intermediate result $\mathbf{Z} = \mathbf{F}'_\mathcal{A} + \mathbf{F}'_\mathcal{B} + \mathbf{F}'_\mathcal{C} = \gamma(\mathbf{X}_a \mathbf{w}_a + \mathbf{X}_b \mathbf{w}_b)$

        \STATE Party $\mathcal{C}$ calculates $\mathbf{u} = q_0 \mathbf{Z}^3$, $\mathbf{v} = q_1 \mathbf{Z}$ and shares $\langle \mathbf{u} \rangle_0$, $\langle \mathbf{v} \rangle_0$ to party $\mathcal{A}$ and $\langle \mathbf{u} \rangle_1$, $\langle \mathbf{v} \rangle_1$ to party $\mathcal{B}$
        
        \STATE Party $\mathcal{A}$ calculates $\mathbf{F}_\mathcal{A} = \gamma^{-3} \langle \mathbf{u} \rangle_0 + \gamma^{-1} \langle \mathbf{v} \rangle_0 + \alpha_0$ and sends $\mathbf{F}_\mathcal{A}$ to party $\mathcal{C}$

        \STATE Party $\mathcal{B}$ calculates $\mathbf{F}_\mathcal{B} = \gamma^{-3} \langle \mathbf{u} \rangle_1 + \gamma^{-1} \langle \mathbf{v} \rangle_1 + \alpha_1 - \mathbf{Y}$ and sends $\mathbf{F}_\mathcal{B}$ to party $\mathcal{C}$

        \STATE Party $\mathcal{C}$ calculates $\mathbf{S} = \mathbf{\hat{Y}} - \mathbf{Y} + \alpha = \mathbf{F}_\mathcal{A} + \mathbf{F}_\mathcal{B} + q_2 \approx \sigma(\mathbf{X}_a \mathbf{w}_a + \mathbf{X}_b \mathbf{w}_b) - \mathbf{Y} + \alpha$ 
        
    \end{algorithmic}
\end{algorithm}

\begin{table}[tb]
    \centering
    \begin{tabular}{|c|c|c|c|}
        \hline
        Party  & Data shares & Weight shares & Random masks \\
        \hline
        $\mathcal{A}$ & $\mathbf{X}_a$, $\langle \mathbf{X}_b \rangle _0$ & $\langle \mathbf{w}_a \rangle _0$, $\langle \mathbf{w}_b \rangle _0$ & $\alpha, \beta, \gamma$      \\
        $\mathcal{B}$ & $\mathbf{X}_b$, $\langle \mathbf{X}_a \rangle _1$ & $\langle \mathbf{w}_a \rangle _1$, $\langle \mathbf{w}_b \rangle _1$ & $\alpha, \beta, \gamma$      \\
        $\mathcal{C}$ & $\langle \gamma \mathbf{X}_a \rangle _0$, $\langle \gamma \mathbf{X}_b \rangle _1$ & $\langle \mathbf{w}_a \rangle _1$, $\langle \mathbf{w}_b \rangle _0$ & $r_0, r_1$     \\
        \hline
    \end{tabular}
    \caption{Data shares after initialization}
    \label{tab:shares}
\end{table}

\begin{algorithm*}[tbh!]
    \caption{Vertical Federated Linear Model via TTP-aided Secret Sharing}
    \label{alg:ttp-vfl}
    \textbf{Input}: Data $\mathbf{X}_a$ hold by party $\mathcal{A}$, data $\mathbf{X}_b$ and  label $\mathbf{Y}$ hold by party $\mathcal{B}$, number of epochs $T$, regression type $\mathcal{F}$ \\
    \textbf{Output}: Trained weights $\mathbf{w}_a$ hold by party $\mathcal{A}$, $\mathbf{w}_b$ hold by party $\mathcal{B}$
    \begin{algorithmic}[1] 

        \STATE \textbf{\underline{Initialize random masks}}

        \STATE Party $\mathcal{A}$ locally generates $T$ random masks $\alpha_0$, $\beta_0$, $\gamma$, sends them to party $\mathcal{B}$

        \STATE Party $\mathcal{B}$ locally generates $T$ random masks $\alpha_1$, $\beta_1$, sends them to party $\mathcal{A}$

        \STATE Party $\mathcal{C}$ locally generates $T$ random masks $r_0, r_1$
        
        \STATE \textbf{\underline{Initialize shared secrets}}

        \STATE Party $\mathcal{A}$ locally generate shares $\langle \mathbf{X}_a \rangle _0$ and $\langle \mathbf{X}_a \rangle _1$, $\langle \mathbf{w}_a \rangle _0$ and $\langle \mathbf{w}_a \rangle _1$,

        \STATE Party $\mathcal{B}$ locally generate shares $\langle \mathbf{X}_b \rangle _0$ and $\langle \mathbf{X}_b \rangle _1$, $\langle \mathbf{w}_b \rangle _0$ and $\langle \mathbf{w}_b \rangle _1$,
        
        \STATE Party $\mathcal{A}$ sends $\langle \mathbf{X}_a \rangle _1$, $\langle \mathbf{w}_a \rangle _1$ to party $\mathcal{B}$, 
        Party $\mathcal{B}$ sends $\langle \mathbf{X}_b \rangle _0$, $\langle \mathbf{w}_b \rangle _0$ to party $\mathcal{A}$

        \IF{$\mathcal{F}$ is Linear Regression}
        \STATE Party $\mathcal{A}$ sends $\langle \mathbf{X}_a \rangle _0$, $\langle \mathbf{w}_a \rangle _1$ to party $\mathcal{C}$, party $\mathcal{B}$ sends $\langle \mathbf{X}_b \rangle _1$, $\langle \mathbf{w}_b \rangle _0$ to party $\mathcal{C}$

        \ELSE
        \STATE Party $\mathcal{A}$ sends $\langle \gamma \mathbf{X}_a \rangle _0$, $\langle \mathbf{w}_a \rangle _1$ to party $\mathcal{C}$, party $\mathcal{B}$ sends $\langle \gamma \mathbf{X}_b \rangle _1$, $\langle \mathbf{w}_b \rangle _0$ to party $\mathcal{C}$
        
        \ENDIF

        \STATE \textbf{\underline{Training:}}

        \FOR{t = 1 to $T$}
        
        \STATE \textbf{\underline{Forward phase}}

        \IF{$\mathcal{F}$ is Linear Regression}
        \STATE Party $\mathcal{A}, \mathcal{B}, \mathcal{C}$ collaboratively calculate the masked sum $\mathbf{S}$ through Algorithm~\ref{alg:lin_fw}
        \ELSE
        \STATE Party $\mathcal{A}, \mathcal{B}, \mathcal{C}$ collaboratively calculate the masked sum $\mathbf{S}$ through Algorithm~\ref{alg:log_fw}
        \ENDIF
        
        \STATE \textbf{\underline{Backward phase}}
        
        \STATE Party $\mathcal{C}$ shares the masked sum $\mathbf{S} = \langle \mathbf{S} \rangle _0 + \langle \mathbf{S} \rangle _1$, sends $\langle \mathbf{S} \rangle _0$ to party $\mathcal{A}$ and $\langle \mathbf{S} \rangle _1$ to party $\mathcal{B}$

        \STATE Party $\mathcal{C}$ calculates $\mathbf{H}_0 = \langle \mathbf{X}_b \rangle_1^T  \langle \mathbf{S} \rangle _0 + r_1$ and sends $\mathbf{H}_0, r_0$ to party $\mathcal{A}$

        \STATE Party $\mathcal{C}$ calculates $\mathbf{H}_1 = \langle \mathbf{X}_a \rangle_0^T  \langle \mathbf{S} \rangle _1 + r_0$ and sends $\mathbf{H}_1, r_1$ to party $\mathcal{B}$

        \STATE Party $\mathcal{A}$ calculates $\langle \mathbf{g}_a \rangle_0 = \mathbf{X}_a^T (\langle \mathbf{S} \rangle_0 - \alpha) - r_0$ and $\langle \mathbf{g}_b \rangle_0 = \mathbf{H}_0 + \langle \mathbf{X}_b \rangle_0^T \langle \mathbf{S} \rangle_0$

        \STATE Party $\mathcal{B}$ calculates $\langle \mathbf{g}_a \rangle_1 = \mathbf{H}_1 + \langle \mathbf{X}_a \rangle_1^T \langle \mathbf{S} \rangle_1$ and $\langle \mathbf{g}_b \rangle_1 = \mathbf{X}_b^T (\langle \mathbf{S} \rangle_1 - \alpha) - r_1$

        \STATE \textbf{\underline{Update weights}}

        \STATE Party $\mathcal{A}$ updates $\langle \mathbf{w}_a \rangle_0$ and $\langle \mathbf{w}_b \rangle_0$ by  $\langle \mathbf{w}_a \rangle_0 = \langle \mathbf{w}_a \rangle_0 - \eta \cdot \langle \mathbf{g}_a \rangle_0 + \beta_0$, $\langle \mathbf{w}_b \rangle_0 = \langle \mathbf{w}_b \rangle_0 - \eta \cdot \langle \mathbf{g}_b \rangle_0 + \beta_1$

        \STATE Party $\mathcal{B}$ updates $\langle \mathbf{w}_a \rangle_1$ and $\langle \mathbf{w}_b \rangle_1$ by  $\langle \mathbf{w}_a \rangle_1 = \langle \mathbf{w}_a \rangle_1 - \eta \cdot \langle \mathbf{g}_a \rangle_1 - \beta_0$, $\langle \mathbf{w}_b \rangle_1 = \langle \mathbf{w}_b \rangle_1 - \eta \cdot \langle \mathbf{g}_b \rangle_1 - \beta_1$

        \STATE \textbf{\underline{Re-share updated weights}}
        
        \STATE Party $\mathcal{A}$ sends $\langle \mathbf{w}_b \rangle_1$ to party $\mathcal{C}$
        \STATE Party $\mathcal{B}$ sends $\langle \mathbf{w}_a \rangle_1$ to party $\mathcal{C}$
        
        \ENDFOR

        \STATE \textbf{\underline{Reconstruct model weights}}

        \STATE Party $\mathcal{A}$ sends $\langle \mathbf{w}_b \rangle_0$ to party $\mathcal{B}$
        \STATE Party $\mathcal{B}$ sends $\langle \mathbf{w}_a \rangle_1$ to party $\mathcal{A}$
        \STATE Party $\mathcal{A}$ reconstructs $\mathbf{w}_a$ = $\langle \mathbf{w}_a \rangle_0 + \langle \mathbf{w}_a \rangle_1$
        \STATE Party $\mathcal{B}$ reconstructs $\mathbf{w}_b$ = $\langle \mathbf{w}_b \rangle_0 + \langle \mathbf{w}_b \rangle_1$
        
    \end{algorithmic}
\end{algorithm*}

\section{Design}

\subsection{Designed Algorithm}

In this section, we propose an algorithm of performing basic machine learning tasks (linear regression and logistic regression) through an efficient TTP-aided additive secret sharing scheme with masking under a semi-honest, non-colluding security settings where the two participants hold vertically-partitioned training data. 

We observed that the secret multiplications occupy the majority of computation and communication overhead when the matrix dimension grows large. In our algorithm, the secrets are shared in 2 parts instead of 3 to avoid the high computation complexity. In addition, our algorithm removes the dependency of Beaver Multiplication Triples for secret multiplications so that the expensive communication overhead of exchanging masked secrets is avoided. Instead, the integrity and privacy of secret multiplications are achieved by offloading the linear components of these computations to the TTP. \\
\textbf{Initialization Phase}
In the initialization phase, each participant shares its data $\mathbf{X}$ and weights $\mathbf{w}$ with the other participants as well as the TTP in the setup phase (See Table~\ref{tab:shares}). The participants and the coordinator also generate corresponding random masks $\alpha, \beta, \gamma$ and $r$ to prevent possible privacy leakage from intermediate calculation results.\\
\textbf{Forward Phase}
In the forward phase, each participant locally computes an intermediate result of the sub-secrets they hold, and the coordinator securely calculates a masked sum $\mathbf{S} = \mathbf{\hat{Y}} - \mathbf{Y} + \alpha$. For linear regression, the participants first mask the local product and then send the masked data to the coordinator, as described in Algorithm~\ref{alg:lin_fw}. Additional steps are required for logistic regression, where the coordinator and the participants interact for two more rounds in order to securely calculate the approximated result of the sigmoid function, as described in Algorithm~\ref{alg:log_fw}. \\
\textbf{Backward Phase}
In the backward phase, the coordinator sends the shared $\mathbf{S}$ as well as the linear components of gradients $\mathbf{H}_0$, $\mathbf{H}_1$ to the participants, allowing them to compute the shared gradients and update corresponding shared weights locally. The shared weights are also shifted by some random values to prevent possible privacy leakage after re-sharing the updated weights with the coordinator.

Detailed process of the proposed algorithm is shown in Algorithm~\ref{alg:ttp-vfl}. We apply similar truncation rule of fix-point multiplications that follows~\cite{7958569}.

\subsection{Security Analysis}




Our proposed method assumes the presence of a trusted third party as the coordinator and two data providers as the participants. We consider a semi-honest and non-colluding adversary model, where different parties strictly follow the protocol but will try their best to infer other parties' confidential data based on their information. We analyze the security of our method with the following two lemmas:
\begin{lemma}
Given an equation $a = x + r$ in ring $\mathbb{Z}_{2^l}$, and $r$ is a uniformly distributed random variable, there exist $2^l$ numbers of $x'$ that can be masked into $a$ (i.e. $a = x' + r' = x + r$) .
\end{lemma}
\noindent \textbf{Proof} Since $r$ is a random number uniformly distributed in $\mathbb{Z}_{2^l}$, there exists in total $2^l$ possible values of $r$. For each possible value of $r$, there exists a $x \in \mathbb{Z}_{2^l}$ such that $x = a - r$ based on the basic property of ring arithmatics.
\begin{lemma}
Given an equation $a = rx$ in ring $\mathbb{Z}_{2^l}$ and $r$ being a uniformly distributed random variable with a multiplicative inverse, there exist $2^{l - 1}$ numbers of $x'$ that can be masked into $a$ (i.e. $a = r'x' = rx$).
\end{lemma}
\noindent \textbf{Proof} Since $r$ is a random number with a multiplicative inverse in $\mathbb{Z}_{2^l}$, $r$ is coprime to $2^l$. Therefore, there are in total $\varphi(2^l)$ possible values of $r$ where $\varphi$ is the Euler quotient function. By the definition of $\varphi$, \[ \varphi(2^l) = 2^l - 2^{l - 1} = 2^{l - 1}\] For each possible value of $r$, there exists a $x \in \mathbb{Z}_{2^l}$ such that $x = r^{-1}a$ based on the basic property of ring arithmatics.

Given a sufficiently large $l$ (e.g. 32, 64), we argue that the probability of guessing the correct value of the secret number $x$ is sufficiently small and the system achieves expected security.\\
\textbf{The TTP can not reconstruct the masked secrets}
\begin{itemize}
    \item In the setup phase, the TTP holds shares of the training data and weights of the participants. Based on Lemma 1, the TTP is unable to reconstruct the correct value of the secret.
    \item In the forward phase, each data holder first computes a masked local intermediate result and sends it to the TTP. 1) For the received data $\mathbf{F}_\mathcal{A}$ and $\mathbf{F}_\mathcal{B}$ from party $\mathcal{A}$ and $\mathcal{B}$, the TTP $\mathcal{C}$ is unable to remove the random mask $\alpha_0$ and $\alpha_1$ to obtain the original result in both linear regression and logistic regression tasks based on Lemma 1 and 2 respectively. 2) For the aggregated result $\mathbf{S} = \mathbf{\hat{Y}} - \mathbf{Y} + \alpha$, the TTP $\mathcal{C}$ is unable to remove the random mask $\alpha$ to obtain the actual sum $\mathbf{S}$ based on Lemma 1.
    \item In the backward gradient update phase, the TTP $\mathcal{C}$ receives no information from the participants and the data remains secure.
    \item In the re-sharing phase, the TTP receives the updated $\langle \mathbf{w}_b \rangle_0$ and $\langle \mathbf{w}_a \rangle_1$. However, since both values are shifted by not only the gradients but also the random masks $\beta_0$, $\beta_1$, the TTP is unable to obtain the updated gradients based on Lemma 1.
\end{itemize}

\noindent \textbf{The participants can not reconstruct the masked secrets}
\begin{itemize}
    \item In the setup phase, the participants only hold shares of the training data and weights of the other participant and negotiated random masks. Based on Lemma 1, the participants are unable to reconstruct the correct value of each other's secret.
    \item In the forward phase, the participants receive no information from other parties and the data remains secure.
    \item In the backward gradient update phase, $\mathcal{A}$ and $\mathcal{B}$ obtain only shares of the aggregated result $\langle \mathbf{S} \rangle_0$ and $\langle \mathbf{S} \rangle_1$ from $\mathcal{C}$ respectively so that none of them can successfully reconstruct the aggregated result $\mathbf{S}$ based on Lemma 1. $\mathcal{A}$ and $\mathcal{B}$ also obtain intermediate results with random masks $r_1$ and $r_0$ respectively so that $\mathcal{A}$ and $\mathcal{B}$ can not reconstruct the original data based on Lemma 1.
    \item In the re-sharing phase, the participants receive no information from other parties and the data remains secure.
\end{itemize}


\begin{table}[tb]
    \centering
    \renewcommand{\arraystretch}{1.25}
    \resizebox{\columnwidth}{!}{\begin{tabular}{|c|c|c|c|c|c|c|c|}
        \hline
        \multicolumn{2}{|c|}{\multirow{2}{*}{Complexity}} & \multicolumn{2}{c|}{Chameleon} & \multicolumn{2}{c|}{ABY3} & \multicolumn{2}{c|}{\projectname}\\
        \cline{3-8}
        \multicolumn{1}{|c}{} & & Comm. & Comp. & Comm. & Comp. & Comm. & Comp. \\
        \hline
        \multirow{3}{*}{\begin{turn}{90}Linear\end{turn}} & Forward & $nd$ & $4nd$ & $2n$ & $6nd$ & $n$ & $2nd$ \\
        \cline{2-8}
        & Backward & $nd$ & $4nd$ & $2d$ & $6nd$ & $2n + 5d$ & $2nd$ \\
        \cline{2-8}
        & Total & $2nd$ & $8nd$ & $2n + 2d$ & $12nd$ & $3n + 5d$ & $4nd$ \\
        \hline
        \multirow{3}{*}{\begin{turn}{90}Logistic\end{turn}} & Forward & $nd$ & $4nd$ & $9n$ & $6nd$ & $4n$ & $2nd$ \\
        \cline{2-8}
        & Backward & $nd$ & $4nd$ & $2d$ & $6nd$ & $2n + 5d$ & $2nd$ \\
        \cline{2-8}
        & Total & $2nd$ & $8nd$ & $9n + 2d$ & $12nd$ & $6n + 5d$ & $4nd$ \\
        \hline
    \end{tabular}}
    \caption{Complexity comparisons in one training epoch between Chameleon, ABY3 and \projectname}
    \label{tab:complexity}
\end{table}

\subsection{Efficiency Analysis}

We analyze the theoretical efficiency of our approach compared with other frameworks. We compare Chameleon which utilizes the Du-Atallah (DA) protocol~\cite{991526} for TTP-aided randomness generation approaches and ABY3~\cite{10.1145/3243734.3243760} for replicated secret sharing schemes. For simplicity, we skip the randomness generation phase of these frameworks which is more costly than ours and focus on analyzing the major cost of matrix multiplications and communications in the execution phase. We demonstrate that our approach achieves much higher efficiency even without this advantage. Denote the batch size as $n$ and the feature dimension as $d$. \\
\textbf{Chameleon}\\
The Chameleon framework adopts the Du-Atallah protocol for offloading the randomness generation process to the TTP. When two parties $\mathcal{A}$ and $\mathcal{B}$ securely compute a secret multiplication $\langle z \rangle = \langle x \rangle \cdot \langle y \rangle$ where $x$, $y$ are hold by $\mathcal{A}$, $\mathcal{B}$ respectively, the TTP first generates a multiplication triple $(a, b, c)$ such that $c = ab$ and sends $(\langle c \rangle_0, a)$ to party $\mathcal{A}$ and $(\langle c \rangle_1, b)$ to party $\mathcal{B}$. In the execution phase, party $\mathcal{A}$ sends $x + a$ to party $\mathcal{B}$ and party $\mathcal{B}$ sends $y - b$ to party $\mathcal{A}$. Then, party $\mathcal{A}$ locally computes the share $\langle z \rangle_0 = x \cdot (y - b) - \langle c \rangle_0$ and party $\mathcal{B}$ locally computes the share $\langle z \rangle_1 = (x + a) \cdot b - \langle c \rangle_1$.

In VFL linear regressions via Chameleon where the participants are data holders, the training data $\mathbf{X}$ can remain clear-text since sharing $\mathbf{X}$ is unnecessary and inefficient. Denote party $\mathcal{P}_0 = a$ and party $\mathcal{P}_1 = b$, then in the forward phase, each party $\mathcal{P}_i$ where $i \in \{0, 1\}$ needs to perform \[
\mathbf{X}_{\mathcal{P}_i} \langle \mathbf{w}_{\mathcal{P}_i} \rangle_i + \langle \mathbf{X}_{\mathcal{P}_i} \langle \mathbf{w}_{\mathcal{P}_i} \rangle_{1 - i} \rangle_i + \langle \mathbf{X}_{\mathcal{P}_{1 - i}} \langle \mathbf{w}_{\mathcal{P}_{1 - i}} \rangle_i \rangle_i
\]
Note that only $\mathbf{X}_{\mathcal{P}_i} \langle \mathbf{w}_{\mathcal{P}_i} \rangle_i$ can be locally computed by party $\mathcal{P}_i$ and the other 2 terms would require extra interactions between the two parties. For $\mathbf{X}_{\mathcal{P}_i} \langle \mathbf{w}_{\mathcal{P}_i} \rangle_{\mathcal{P}_{1 - i}}$, $\mathcal{P}_i$ needs to mask and send $\mathbf{X}_{\mathcal{P}_i}$ to $\mathcal{P}_{1 - i}$, bringing a communication cost of $O(nd)$. Similarly for $\mathbf{X}_{\mathcal{P}_{1 - i}} \langle \mathbf{w}_{\mathcal{P}_{1 - i}} \rangle_i$, $\mathcal{P}_i$ needs to mask and send $\langle \mathbf{w}_{\mathcal{P}_{1 - i}} \rangle_i $ which brings another $O(d)$ communication cost. The parties then perform local computations with the received data. In the backward phase, denote $\mathbf{A} = \hat{\mathbf{Y}} - \mathbf{Y}$, then each party needs to perform \[
    \mathbf{X}_{\mathcal{P}_i}^{T} \langle \mathbf{A}_{\mathcal{P}_i} \rangle_i + \langle \mathbf{X}_{\mathcal{P}_i}^{T} \langle \mathbf{A}_{\mathcal{P}_i} \rangle_{1 - i} \rangle_i + \langle \mathbf{X}_{\mathcal{P}_{1 - i}}^{T} \langle \mathbf{A}_{\mathcal{P}_{1 - i}} \rangle_i \rangle_i
\]
which takes similar communication and computation costs as the forward phase. The total cost for communication and computation would be $O(2nd)$ and $O(8nd)$ respectively for one training epoch in the linear regression scenario.

As Chameleon provides few details on logistic regression, we simulate its best-case scenario with minimax approximation. Denote $\mathbf{Z} = \mathbf{X}\mathbf{w}$, the backward phase remains the same as the linear regression case. The forward phase would require a few more computations on $\mathbf{Z}^3$ and $\mathbf{Z}^2$ which only takes $O(n)$ complexity and is trivial compared with the total cost. Compared with linear regression, the additional cost of logistic regression is negligible. We therefore consider the cost of logistic regression to be roughly the same as that of linear regression.\\
\textbf{ABY3}\\
For replicated secret sharing schemes such as ABY3, all secrets need to be shared in 3 parts for secure computations where each party holds 2 shares of a secret. For each shared secret multiplication $\langle z \rangle = \langle x \rangle \cdot \langle y \rangle = \langle x_0 + x_1 + x_2 \rangle \cdot \langle y_0 + y_1 + y_2 \rangle$, each party performs 3 local multiplications $z_0 = x_0 y_0 + x_0 y_1 + x_1 y_0 + \alpha_0$, $z_1 = x_1 y_1 + x_1 y_2 + x_2 y_1 + \alpha_1$, and $z_2 = x_2 y_2 + x_2 y_0 + x_0 y_2 + \alpha_2$ where $\alpha$ is the pre-generated randomness such that $\alpha_0 + \alpha_1 + \alpha_2 = 0$. Each party $i$ then sends $z_i$ to the party $i - 1$ so that the 2-out-of-3 secret replication is preserved.\\
This sharing mechanism is incompatible with one clear-text operand and takes nearly twice more computation cost in the two-party VFL setting, as each party needs to perform two secret multiplications to get the intermediate result instead of one. Specifically, each party $\mathcal{P}_i$ where $i \in \{0, 1, 2\}$ in the forward phase in linear regression needs to perform both 
\[
    \langle \mathbf{X}_{\mathcal{A}} \rangle_i \langle \mathbf{w}_{\mathcal{A}} \rangle_i + \langle \mathbf{X}_{\mathcal{A}} \rangle_i \langle \mathbf{w}_{\mathcal{A}} \rangle_{i + 1} + \langle \mathbf{X}_{\mathcal{A}} \rangle_{i + 1} \langle \mathbf{w}_{\mathcal{A}} \rangle_{i} + \alpha_i
\]
and \[
    \langle \mathbf{X}_{\mathcal{B}} \rangle_i \langle \mathbf{w}_{\mathcal{B}} \rangle_i + \langle \mathbf{X}_{\mathcal{B}} \rangle_i \langle \mathbf{w}_{\mathcal{B}} \rangle_{i + 1} + \langle \mathbf{X}_{\mathcal{B}} \rangle_{i + 1} \langle \mathbf{w}_{\mathcal{B}} \rangle_{i} + \beta_i
\] where $i + 1$ is mod 3 and $\alpha$, $\beta$ are pre-generated randomness. Similar problem exists in the backward process, bringing a total computation cost of $O(12nd)$ in one linear regression training epoch. The communication cost is rather small compared with Chameleon, as ABY3 only needs to re-share the multiplication result with another party. This yields a total communication cost of $O(2n + 2d)$ in one linear regression training epoch. 

For logistic regression, ABY3 follows the approach of ~\cite{7958569} which replaces the sigmoid function with a piece-wise approximation evaluated via GC. According to~\cite{7958569}, this process takes 3 extra OTs and 1 GC execution, bringing an additional communication cost of around $O(7n)$ and negligible computation cost. \\
\textbf{\projectname}\\
Our approach takes the advantage of collaborative computations as well as an efficient secret sharing scheme with masking in order to achieve the best performance. According to algorithm~\ref{alg:ttp-vfl},  the forward phase of linear regression takes a computation cost of $O(2nd)$ as each party locally computes 2 sub-secret multiplications and a communication cost of $O(n)$. Similarly, the backward and re-sharing phase of linear regression take another computation cost of $O(2nd)$ and communication cost of $O(2n + 5d)$.

For logistic regression, the complexity of the backward and re-sharing phase remains the same, while the forward phase takes an additional $O(3n)$ communication overhead and negligible computation overhead.

\begin{figure*}[hbt!]
\includegraphics[width=\textwidth]{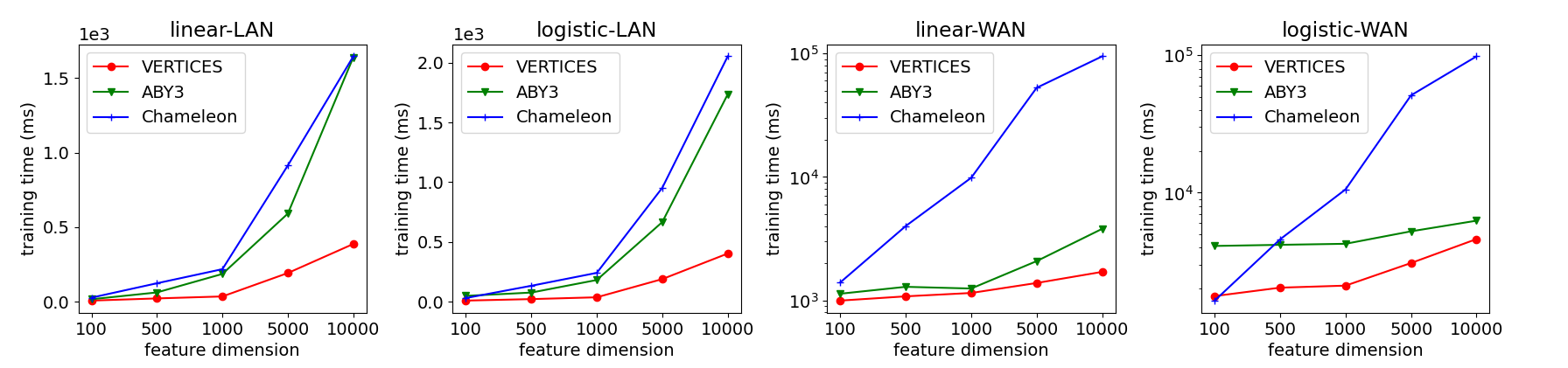} 
\caption{Training time comparisons between different feature dimensions at batch size = 512}
\label{fig:time-feautre}
\end{figure*}

\begin{figure*}[hbt!]
\includegraphics[width=\textwidth]{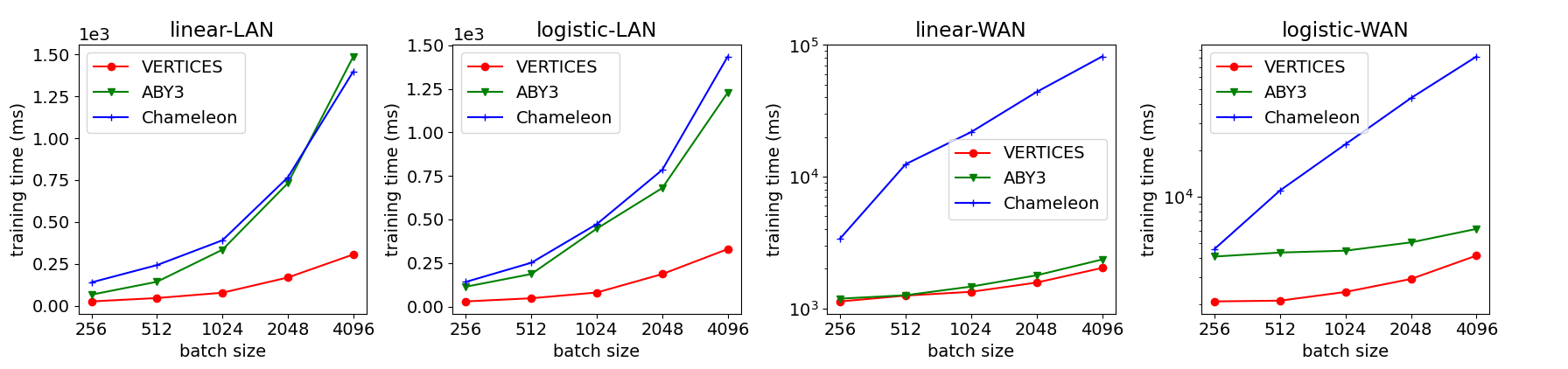} 
\caption{Training time comparisons between different batch size at feature dimension = 1000}
\label{fig:time-batch}
\end{figure*}

\begin{figure*}[hbt!]
\includegraphics[width=\textwidth]{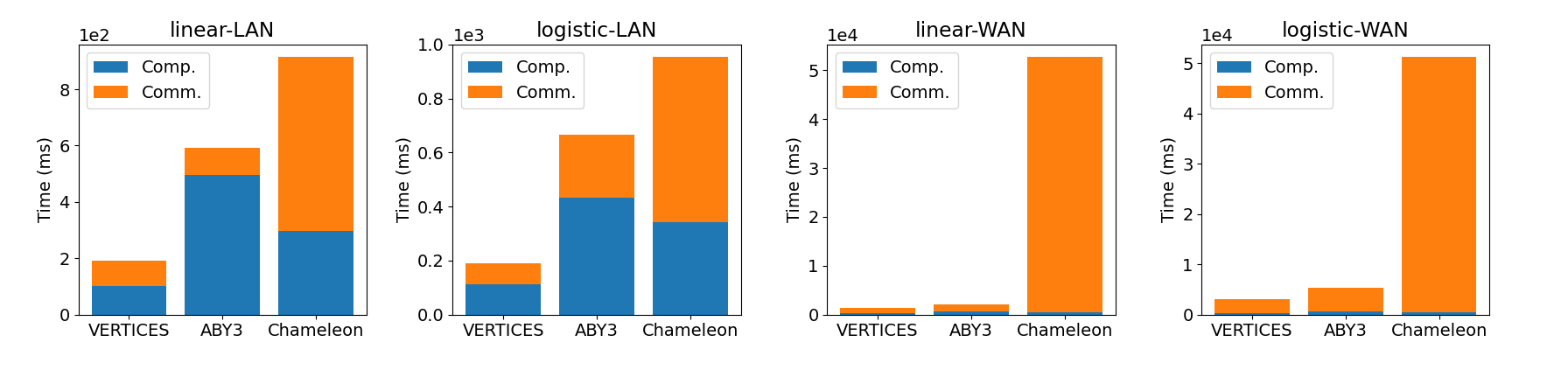}
\caption{Cost decomposition between different frameworks}
\label{fig:cost-decomposition}
\end{figure*}

\section{Experiment}

We evaluated the accuracy and efficiency of the proposed algorithm compared with the other existing solutions. We implemented our algorithm in C++ with third library dependencies of EMP-tools~\cite{emp-toolkit} and Eigen~\cite{eigenweb} and tested its performance on datasets with different dimensions. 
\subsection{Experiment Setup}
We tested our implementations on 3 servers with Intel(R) Xeon(R) Gold 5115 CPU at 2.40GHz and 128G RAM in a local area network (LAN) with inter-node latency $< 0.1$ ms and simulated the wide area network (WAN) environment by increasing the inter-node latency to 50 ms. 
We tested on both different feature dimensions as well as different batch sizes and trained the model for 10 epochs. The features are evenly partitioned into two parts hold by the two parties $\mathcal{A}$ and $\mathcal{B}$ respectively, where labels are all hold by party $\mathcal{B}$.

\begin{table}[tb]
    \centering
    \begin{tabular}{|c|c|c|}
        \hline
        Name & $\#$ Samples (selected) & $\#$ Features \\
        \hline
        MNIST & 14780 & 784 \\
        \hline
        Citeseer & 1369 & 3703\\
        \hline
    \end{tabular}
    \caption{Brief description of data dimensions from experiment datasets}
    \label{tab:datasets}
\end{table}

\begin{table}[tb]
    \centering
    \begin{tabular}{|c|c|c|c|c|}
        \hline
        \multirow{2}{*}{Accuracy} & \multicolumn{2}{c|}{Logistic} & \multicolumn{2}{c|}{\projectname} \\
        \cline{2-5}
        \multicolumn{1}{|c|}{} & Train & Test & Train & Test \\
        \hline
        MNIST & 99.93 & 99.95 & 99.87 & 99.95 \\
        \hline
        Citeseer & 92.33 & 88.67 & 93.33 & 86.13\\
        \hline
    \end{tabular}
    \caption{Accuracy (\%) comparisons between standard logistic regression and our approach in binary classification}
    \label{tab:accuracy}
\end{table}

\subsection{Datasets}
Since our work aims at improving the performance of high-dimensional vertically-partitioned learning, we emphasize the speed up of our algorithm under different data scales. To better demonstrate the effectiveness of our approach, we use synthesized dataset so that the number of features and samples can be conveniently controlled as an experiment variable. For correctness and accuracy guarantees, we also tested our algorithm on datasets MNIST~\cite{lecun-mnisthandwrittendigit-2010} and Citeseer~\cite{10.1145/276675.276685} for logistic regression tasks. 
\subsection{Experiment Result}
\textbf{End-to-end Efficiency} 
As Figure~\ref{fig:time-feautre} and \ref{fig:time-batch} have shown, our proposed algorithm can achieve a much shorter training time compared with the existing approaches under vertically partitioned high-dimensional machine learning settings in LAN settings. Under WAN settings where we assume a network propagation delay of around 50 ms, our algorithm achieves a similar performance with ABY3 mainly due to the fact that the network delay occupies nearly all of the total overhead to the point that our improvements on computations have very little effects on the overall training time, as Figure~\ref{fig:cost-decomposition} has shown. However, our approach still outperforms the Chameleon framework due to its high communication complexity. 

In linear regression tasks under LAN settings, our algorithm reaches between 2.3$\times$ and 5.3$\times$ speed up compared with ABY3's implementation and between 3.8$\times$ and 6.3$\times$ speed up compared with the Chameleon framework. In logistic regression tasks under LAN settings, our algorithm reaches between 3.5$\times$ and 5.3$\times$ speed up compared with ABY3 and between 3.3$\times$ and 6.6$\times$ speed up compared with Chameleon. In addition, both experiments did not include the offline preparation time of ABY3 and Chameleon, where they generate and share certain randomness within the participants. The actual overall execution acceleration is therefore even better than the existing result.\\
\textbf{Accuracy}
We also evaluated how our approximation of the sigmoid function in logistic regression affects the accuracy of the trained model. For simplicity, we extracted data with label 0 (6903 samples) and 1 (7877 samples) from the MNIST dataset, class 2 (668 samples) and 3 (701 samples) from the Citeseer dataset and tested the performance of binary classification on these selected samples. We run a total of 100 training epochs with a learning rate at 0.05, batch size at 128, and train-test split ratio of 0.8. As shown in Table~\ref{tab:accuracy}, the result demonstrates that our approach only experience a small accuracy drop within 2\% in the chosen datasets. 

\section{Related works}

Secure-ML~\cite{7958569} proposed a privacy-preserving machine learning framework that enables two party to securely perform machine learning tasks via secret sharing. However, such 2PC secret sharing schemes suffer from great computation and communication overhead when performing multiplications between shared secrets, as each multiplication would consume a Beaver Multiplication Triple (MT) generated via either HE or OT in the preparation phase to prevent privacy leakage. In machine learning tasks with high-dimensional data where large-scale matrix multiplications are ubiquitous, this offline preparation phase can take orders of magnitude longer time than the online execution phase.

CAESAR~\cite{DBLP:journals/corr/abs-2008-08753} proposed another HE-MPC combined protocol that takes the advantages of both HE and MPC, achieving relatively lower overhead than Secure ML while providing provably-secure privacy guarantees. In this approach, secret multiplications are protected by HE and no longer require pre-generated multiplication triples. Nevertheless, its protocol consists of a large number of encryption and decryption operations of HE in a single training iteration. The computation overhead of HE thus becomes a major part in the overall cost that greatly hinders performance.

\cite{991526} proposed a method introducing a TTP for multiplication triple generation and distribution in secret sharing and has been extensively used in many frameworks such as ShareMind~\cite{10.1007/978-3-540-88313-5_13}, Chameleon~\cite{10.1145/3196494.3196522} and~\cite{gascon2016privacy}. This approach alleviates the pressure of the offline preparation phase by offloading the MT generation task to a trusted third party. It enables the trusted third party to locally generate the multiplication triples and distribute them to the participants, so that the MT generation procedure no longer requires the expensive OT or HE operations between the two secret holders. As a result, the Du-Atallah's protocol significantly reduces the time of offline randomness generation process and therefore greatly improves the overall efficiency of secret-sharing based machine learning.

Other secret sharing schemes such as \cite{10.1145/2976749.2978331} \cite{10.1145/3243734.3243760} \cite{wagh2021f} utilize the replicated sharing technique so that secret multiplications can be done without consuming any beaver triples. In these efficient three-party computation protocols, all secrets are shared in three parts where each party holds two shares of a secret. It takes the unique advantage of such 2-out-of-3 replicated sharing feature and removes the original dependencies on MTs for secret multiplications. Instead, each party obtains the shared result of a secret multiplication by computing the shared secrets locally and then re-share the locally computed result.

\section{Conclusions}

\textbf{Summary}\\
This paper proposed a novel SS-based approach for two-party vertical federated linear models through TTP-aided secret sharing. The suggested method 1) removes the dependency of MT required for secret multiplications in standard secret sharing and 2) utilizes an efficient secret sharing scheme with masks to achieve low computation and communication overhead compared with the common practices. The experiment result demonstrates sufficient efficiency improvements in vertical federated linear model training under LAN and WAN environments. \\
\textbf{Future work}\\
Although this work focused on the designed approach for VFL in two-party settings, more complicated scenarios where multiple data holders participating in the VFL training have not yet been discussed thoroughly in this work. It is expected that future work can take the advantages of the current approach and design a more efficient scheme for multi-party VFL settings.







\bibliographystyle{named}
\bibliography{ijcai23}

\begin{thebibliography}{}

\bibitem[\protect\citeauthoryear{Araki \bgroup \em et al.\egroup
  }{2016}]{10.1145/2976749.2978331}
Toshinori Araki, Jun Furukawa, Yehuda Lindell, Ariel Nof, and Kazuma Ohara.
\newblock High-throughput semi-honest secure three-party computation with an
  honest majority.
\newblock In {\em Proceedings of the 2016 ACM SIGSAC Conference on Computer and
  Communications Security}, CCS '16, page 805–817, New York, NY, USA, 2016.
  Association for Computing Machinery.

\bibitem[\protect\citeauthoryear{Bogdanov \bgroup \em et al.\egroup
  }{2008}]{10.1007/978-3-540-88313-5_13}
Dan Bogdanov, Sven Laur, and Jan Willemson.
\newblock Sharemind: A framework for fast privacy-preserving computations.
\newblock In Sushil Jajodia and Javier Lopez, editors, {\em Computer Security -
  ESORICS 2008}, pages 192--206, Berlin, Heidelberg, 2008. Springer Berlin
  Heidelberg.

\bibitem[\protect\citeauthoryear{Chen \bgroup \em et al.\egroup
  }{2018}]{chen2018logistic}
Hao Chen, Ran Gilad-Bachrach, Kyoohyung Han, Zhicong Huang, Amir Jalali, Kim
  Laine, and Kristin Lauter.
\newblock Logistic regression over encrypted data from fully homomorphic
  encryption.
\newblock {\em BMC medical genomics}, 11:3--12, 2018.

\bibitem[\protect\citeauthoryear{Chen \bgroup \em et al.\egroup
  }{2020}]{DBLP:journals/corr/abs-2008-08753}
Chaochao Chen, Jun Zhou, Li~Wang, Xibin Wu, Wenjing Fang, Jin Tan, Lei Wang,
  Xiaoxi Ji, Alex Liu, Hao Wang, and Cheng Hong.
\newblock When homomorphic encryption marries secret sharing: Secure
  large-scale sparse logistic regression and applications in risk control.
\newblock {\em CoRR}, abs/2008.08753, 2020.

\bibitem[\protect\citeauthoryear{Du and Atallah}{2001}]{991526}
Wenliang Du and M.J. Atallah.
\newblock Privacy-preserving cooperative statistical analysis.
\newblock In {\em Seventeenth Annual Computer Security Applications
  Conference}, pages 102--110, 2001.

\bibitem[\protect\citeauthoryear{Dwork}{2008}]{10.1007/978-3-540-79228-4_1}
Cynthia Dwork.
\newblock Differential privacy: A survey of results.
\newblock In Manindra Agrawal, Dingzhu Du, Zhenhua Duan, and Angsheng Li,
  editors, {\em Theory and Applications of Models of Computation}, pages 1--19,
  Berlin, Heidelberg, 2008. Springer Berlin Heidelberg.

\bibitem[\protect\citeauthoryear{Gasc{\'o}n \bgroup \em et al.\egroup
  }{2016}]{gascon2016privacy}
Adri{\`a} Gasc{\'o}n, Phillipp Schoppmann, Borja Balle, Mariana Raykova, Jack
  Doerner, Samee Zahur, and David Evans.
\newblock Privacy-preserving distributed linear regression on high-dimensional
  data.
\newblock {\em Cryptology ePrint Archive}, 2016.

\bibitem[\protect\citeauthoryear{Giles \bgroup \em et al.\egroup
  }{1998}]{10.1145/276675.276685}
C.~Lee Giles, Kurt~D. Bollacker, and Steve Lawrence.
\newblock Citeseer: An automatic citation indexing system.
\newblock In {\em Proceedings of the Third ACM Conference on Digital
  Libraries}, DL '98, page 89–98, New York, NY, USA, 1998. Association for
  Computing Machinery.

\bibitem[\protect\citeauthoryear{Guennebaud \bgroup \em et al.\egroup
  }{2010}]{eigenweb}
Ga\"{e}l Guennebaud, Beno\^{i}t Jacob, et~al.
\newblock Eigen v3.
\newblock http://eigen.tuxfamily.org, 2010.

\bibitem[\protect\citeauthoryear{Kim \bgroup \em et al.\egroup
  }{2018}]{kim2018secure}
Miran Kim, Yongsoo Song, Shuang Wang, Yuhou Xia, Xiaoqian Jiang, et~al.
\newblock Secure logistic regression based on homomorphic encryption: Design
  and evaluation.
\newblock {\em JMIR medical informatics}, 6(2):e8805, 2018.

\bibitem[\protect\citeauthoryear{LeCun and
  Cortes}{2010}]{lecun-mnisthandwrittendigit-2010}
Yann LeCun and Corinna Cortes.
\newblock {MNIST} handwritten digit database.
\newblock 2010.

\bibitem[\protect\citeauthoryear{Liu \bgroup \em et al.\egroup
  }{2022}]{liu2022vertical}
Yang Liu, Yan Kang, Tianyuan Zou, Yanhong Pu, Yuanqin He, Xiaozhou Ye,
  Ye~Ouyang, Ya-Qin Zhang, and Qiang Yang.
\newblock Vertical federated learning, 2022.

\bibitem[\protect\citeauthoryear{Mohassel and
  Rindal}{2018}]{10.1145/3243734.3243760}
Payman Mohassel and Peter Rindal.
\newblock Aby3: A mixed protocol framework for machine learning.
\newblock In {\em Proceedings of the 2018 ACM SIGSAC Conference on Computer and
  Communications Security}, CCS '18, page 35–52, New York, NY, USA, 2018.
  Association for Computing Machinery.

\bibitem[\protect\citeauthoryear{Mohassel and Zhang}{2017}]{7958569}
Payman Mohassel and Yupeng Zhang.
\newblock Secureml: A system for scalable privacy-preserving machine learning.
\newblock In {\em 2017 IEEE Symposium on Security and Privacy (SP)}, pages
  19--38, 2017.

\bibitem[\protect\citeauthoryear{Pinkas \bgroup \em et al.\egroup
  }{2014}]{pinkas2014faster}
Benny Pinkas, Thomas Schneider, and Michael Zohner.
\newblock Faster private set intersection based on $\{$OT$\}$ extension.
\newblock In {\em 23rd $\{$USENIX$\}$ Security Symposium ($\{$USENIX$\}$
  Security 14)}, pages 797--812, 2014.

\bibitem[\protect\citeauthoryear{Riazi \bgroup \em et al.\egroup
  }{2018}]{10.1145/3196494.3196522}
M.~Sadegh Riazi, Christian Weinert, Oleksandr Tkachenko, Ebrahim~M. Songhori,
  Thomas Schneider, and Farinaz Koushanfar.
\newblock Chameleon: A hybrid secure computation framework for machine learning
  applications.
\newblock In {\em Proceedings of the 2018 on Asia Conference on Computer and
  Communications Security}, ASIACCS '18, page 707–721, New York, NY, USA,
  2018. Association for Computing Machinery.

\bibitem[\protect\citeauthoryear{Rivest and Dertouzos}{1978}]{Rivest1978ONDB}
Ronald~L. Rivest and Michael~L. Dertouzos.
\newblock On data banks and privacy homomorphisms.
\newblock 1978.

\bibitem[\protect\citeauthoryear{Wagh \bgroup \em et al.\egroup
  }{2021}]{wagh2021f}
Sameer Wagh, Shruti Tople, Fabrice Benhamouda, Eyal Kushilevitz, Prateek
  Mittal, and Tal Rabin.
\newblock F: Honest-majority maliciously secure framework for private deep
  learning.
\newblock {\em Proceedings on Privacy Enhancing Technologies},
  2021(1):188--208, 2021.

\bibitem[\protect\citeauthoryear{Wang \bgroup \em et al.\egroup
  }{2016}]{emp-toolkit}
Xiao Wang, Alex~J. Malozemoff, and Jonathan Katz.
\newblock {EMP-toolkit: Efficient MultiParty computation toolkit}.
\newblock \url{https://github.com/emp-toolkit}, 2016.

\bibitem[\protect\citeauthoryear{Yang \bgroup \em et al.\egroup
  }{2019}]{DBLP:journals/corr/abs-1902-04885}
Qiang Yang, Yang Liu, Tianjian Chen, and Yongxin Tong.
\newblock Federated machine learning: Concept and applications.
\newblock {\em CoRR}, abs/1902.04885, 2019.

\bibitem[\protect\citeauthoryear{Yang \bgroup \em et al.\egroup
  }{2023}]{yang2023survey}
Liu Yang, Di~Chai, Junxue Zhang, Yilun Jin, Leye Wang, Hao Liu, Han Tian, Qian
  Xu, and Kai Chen.
\newblock A survey on vertical federated learning: From a layered perspective.
\newblock {\em arXiv preprint arXiv:2304.01829}, 2023.

\end{thebibliography}

\end{document}